# The Scientific Prize Network Predicts Who Pushes the Boundaries of Science

Authors: Yifang Ma[1,2] and Brian Uzzi[1,2*]

[1]Northwestern Institute on Complex Systems (NICO), Northwestern University, Evanston IL, 60208, USA

[2]Kellogg School of Management, Northwestern University, Evanston IL, 60208, USA

Correspondence to: *uzzi@northwestern.edu

**Abstract**: Scientific prizes are among the greatest recognitions a scientist receives from their peers and arguably shape the direction of a field by conferring credibility to persons, ideas, and disciplines, providing financial rewards, and promoting rituals that reinforce scientific communities.  Though debates continue about prizes' overall benefits to science, prizes in science have proliferated.  Thousands of prizes are now awarded across diverse sciences, topics, regions of the world, and for different levels of scientific discovery.  The proliferation of prizes and links among prizes suggest that the prize network embodies information about scientists and ideas poised to grow in acclaim.  Using comprehensive new data on prizes and prizewinners worldwide and across disciplines, we examine the growth dynamics and interlocking relationships found in the worldwide scientific prize network.  We focus on understanding how the knowledge linkages among prizes and scientists' propensities for prizewinning are related to knowledge pathways across disciplines and stratification within disciplines.  Our data cover more than 3,000 different scientific prizes in diverse disciplines and the career histories of 10,455 prizewinners worldwide for more than 100 years.  We find several key links between prizes and scientific advances.  First, despite a proliferation of diverse prizes over time and across the globe, prizes are more concentrated within a relatively small group of scientific elites, and ties within the elites are more clustered, suggesting that a relatively constrained number of ideas and scholars lead science.  For example, 64.1% of prizewinners have won two prizes and 13.7% have one 5 or more prizes.  Second, we find that certain prizes are strongly interlocked within and between disciplines by scientists who win multiple prizes, revealing the key pathways by which knowledge systematically gains credit and spreads through the network.  Third, we find that genealogical and co-authorship networks strongly predict who wins one or more prizes and explains the high level of interconnectedness among acclaimed scientists and their path breaking ideas.

## Introduction

Scientific prizes arguably play many roles in advancing scientific discoveries, yet they have been subjected to few quantitative analyses.  They are among the highest forms of recognition scientists accord one another (1). By promoting high risk, high return science (2-4) or new lines of research (5, 6), prizes identify top scientific achievements (7-10). The Nobel Prize for example, is awarded for work that provides "the greatest benefit to mankind" (11). Prizes have cultural functions. They identify successful role models of who inspire a sense of achievement previously thought to be impossible (8, 11, 12) and act as signals of scientific credibility (13). Many scientists can name of their prizewinning "heroes" and with ritualistic fanfare follow each year's announcement of prizewinners (14, 15). At the University of Chicago, faculty follow a cultural practice of standing and applauding Noble prizewinners when they enter a room. Prizes may also forecast the direction of future scientific investments. Prizewinning papers are turned into patented technologies faster than other similarly cited papers (7, 16) and often include direct or indirectly capital (e.g., Howard Hughes Medical Research Award) that stimulates research (4, 6, 17).

While debate about the positive role of prizes remain unsettled (18-21), scientific prizes continue to proliferate. Thousands of prizes now exist across diverse sciences, topics, regions of the world, and for different levels of scientific discovery (8). Besides the Nobel prize, for example, there are the highly prestigious Albert Einstein Award, the Lasker Award in biomedical science, the Davy Medal in chemistry, Fields Medal in mathematics, and Copley Medal, which is discipline non-specific. While prizes aim to expand attention to scientific ideas, they also connect science. In some cases, a single scientist can be a winner of multiple prizes within and across disciplines. Rainer Weiss, the 2017 Nobel prizewinner in Physics was already an Einstein, Shaw, and Harvey prizewinner. The Clark Medal, a.k.a the "Baby Nobel", has 12 of 23 (52%) medalists who went onto become Nobelists (22, 23). The increasing proliferation of prizes and connections among prizes suggest that the prize network embodies information about scientists poised to grow in acclaim as well as the scientific knowledge likely to propagate within and between fields.

Harriet Zuckerman's landmark work (11) provides a foundation for analysis of the prize network and prizewinners. She studied the similarities and differences in Nobelists' demographic, family, religious, co-authorship, and research topics. With the unprecedented expansion of science and the availability of large scale datasets on prizes, prizewinners, citations, and collaborations worldwide, analyses of the global scientific prize network's properties and its potential association with scientific advances and stratification are now possible. Here, we address two main questions. First, we derive the statistical properties of the global scientific prize network in terms of growth dynamics over time and the ensuing transition probabilities between prizes that can interlock subfields within a disciplines and fields between disciplines. Second, we examine how genealogical ties and co-author ties are associated with prizewinning and the concentration of prizes among scientists. We demonstrate that despite a proliferation of diverse prizes over time and across the globe, prizes are more concentrated within a relatively small scientific elite, and ties within the elite are more clustered, suggesting that a relatively constrained number of ideas and scholars lead science.

For our analysis, we collected comprehensive data on approximately all 3,062 scientific prizes worldwide, which includes more than 10,455 winners spanning more than 100 years in more than 50 countries. For each winner, we recorded their publication and citation data, institutional affiliations, years of scientific activity, genealogical relationships, and co-authorship ties. With these data, we examined fundamental questions about the global scientific prize network. The Materials and Methods section provides detailed information on the data's collection, validation, variable definitions, and descriptive statistics.

Our analysis proceeds as follows. First, we derived the statistical properties of the global scientific prize network, focusing on growth dynamics and the ensuing transition probabilities that reveal how prizes interlock knowledge within and between disciplines. Second, we model how a prizewinner's genealogical and co-authorship networks are associated with prizewinning and the concentration of prizes among a few prizewinners.

**Results**

Fig. 1A shows the sharp annual increase in the number of prizes relative to the growth of scientific sub-disciplines. We observe that the number of prizes grew at a faster rate than the number of new disciplines (per the Web of Science (24)). Up to 1980, there were fewer prizes than there were scientific fields; After 1980, the pattern reverses when prizes outnumber disciplines by almost a 2:1 margin. On per annum basis, the number of scientific prizes has roughly doubled every 20 years; currently there are more than 350 prizes conferred per year compared to about twenty prizes roughly 100 years ago.

The proliferation suggests that opportunities to recognize a greater diversity of ideas and scholars have expanded. However, while the absolute number of prizewinners has grown slowly with time (Fig. 1B), we observe that prizewinning has also become more concentrated within a small set of prizewinners. The pie of prizes expanded, but a smaller group of scholars received bigger slices. The emergence of a subset of scientists who win numerous prizes creates this concentration. This uneven distribution of prizes per scientist is evident along several measures shown in Fig. 2. Fig. 2 plots the number of prizes per prizewinner for 4 dominant disciplines and all prizewinners in our sample. The prize-per-scientist distribution is noteworthy in that it is roughly equivalent across diverse disciplines and in each discipline follows an exponential distribution (with $p<10^{-4}$ compared to a power-law distribution) (25). In modern science, the distribution indicates that fully 64.1% of winners won at least two different prizes over their career, 13.7% of the winners won 5 or more prizes, and some scientists are awarded more than 20 prizes over their careers. Fig. 2 inset shows that the contemporary concentration of prizes in the hands of a few prizewinners is a reverse of the distribution that existed before 1985. Before 1985, the prizes are more evenly distributed among prizewinners than in the post-1985 periods. After 1985, there has been a significant increase in concentration at nearly every level of multiple prize winner ($p<10^{-21}$). These data suggest that the explosive expansion of the prize network has expanded the pie for recognizing acclaimed work and the number of influential scientists in absolute sense, yet, at the same time the science has become more stratified within a relatively small and concentrated scientific elite (1). In the next section, we find that the scientific elite has also become transdisciplinary; there has emerged a concentration of scientists who win multiple prizes not just within one discipline but across different disciplines.

The statistical structure of the prize network further reveals an underlying stratification of acclaim in science, as well as interlocking knowledge pathways between and among prizes. To create a prize network, we defined the nodes as prizes. (e.g., Nobel or Fields Medals) with node size representing a prize's prestige based on average Wikipedia page views per month (26). Links between nodes $i$ and $j$ occur when prizes $i$ and $j$ s are won by the same scientist, $k$. These links are weighted by the count of prizewinners that have won prizes $i$ and $j$.

Fig. 3 shows that the number of scientific prizes varies by discipline and is clustered by discipline (a modularity value of 0.492) (27, 28). Notably, each discipline has a similar status hierarchy of prizes. We ranked prize status in bins of at least 10,000, 1,000, 100, or 10 Wikipedia views. Notably, each discipline has 1-3 elite prizes of extremely high significance (10,000 views), a middle layer of prizes that are of high prestige (1,000 views), and a ring of specialized prizes (10 or 100 views). This pattern of recognition indicates that despite differences in talent, number of areas of specialty, funding, age of field, and number of journals, fields tend to characterize pecking order differences in research quality and confer acclaim within the same tiered stratification system.

Fig. 3 also reveals the interlocks among certain prizes that are formed when prize $i$ and $j$ are won by the same scientist. The interlocking pathways between prizes provide a global picture of the backbone of the network of relationships between prizes and disciplines (29-32). A critical measure of the relationship between prizes and disciplines is encapsulated in the probability of transitioning between being a winner of one prize and to a prizewinner of another prize and so on. The stronger the transition probability, the more likely it is that the ideas of the scientist spread throughout the network. Winning one prize increases a scientist's propensity to win another particular prize.

To quantify the interlock transition probabilities, we formulated a probability adjacency matrix of the prize network. In the matrix, the values $(i, j)$ denote the probability that prize $i$ is a precursor to prize $j$ and vice versa. In our data, the scale of interlocking is broad, with $i$ and $j$ prize pairings having between 1 to 150 dual winners (per Fig. 2). The blocks within the matrix indicate the intensity levels of the transition probabilities between any two prizes.

Fig. 4 shows the transition matrix and indicates three important results. First, prize interlocks are strongest within disciplines. This finding quantitatively reinforces the modularity shown in Fig. 3 and suggests that the strongest relationships among prizes occur between sub-disciplines within the same discipline. Nevertheless, disciplines vary in the interlocking of their prizes. Chemistry and physics are the most interlocked, while and math and general prizes are least. Second, while some prizes appear to be stepping-stones to more prestigious prizes, such as the Clark Medal, being the "Baby Nobel," in most cases high transition probabilities between prizes are similar in both directions. This means that acclaimed ideas can originate in multiple ways within science's prize network. Third, we observe that interlocks occur frequently between disciplines (33). This suggests that the acclaim of these prizewinners and their ideas spread widely throughout the broader prize network (34). For example, (4) found that Howard Hughes Medicine Institute Award winners had their work more highly cited across disciplines and tended to combine more novel lines of inquiry than a control group of similarly-accomplished NIH-funded.

The concentration of interlocks raises questions of what factors are associated with increases in the probability of becoming a multi-prizewinner. Research hypotheses that social and genealogical networks could play a role in prizewinning through genealogy and teamwork (11), which can help impart tacit knowledge, aid in learning and information search (35, 36), and provide access to novel, interdisciplinary ideas (37-39).

To investigate this hypothesis, we linked 2,034 winners to their scientific genealogy trees, and extracted their co-authorship network from publication lists. Two winners connect to one another when they have a genealogy relationship (PhD or Postdoc) or are coauthors on at least one paper (see Materials and Methods). Fig. 5 shows the backbone of the prizewinner network and reveals three important findings. First, there is a giant component of winners connected directly through their social ties. 1,504 (74%) of winners are connected in the giant component when we consider both genealogy ties and co-authorship ties. Second, prizewinners are not randomly distributed throughout science. Rather, they are highly interconnected with one another through intensive research collaborations of genealogy and co-authorship. Notably genealogy ties are relatively evenly dispersed while co-authorship ties are densely packed in the center of the prizewinner's network. Given that a scholar's primary, formative relationship is with their advisor, it suggests that genealogy ties are important to winning at least one prize but winning multiple prizes is associated with expanding one's network to prizewinning co-authors. Third, the interconnections among prizewinners are not localized but span the globe, institutions, and time periods. Thus, not only are prizes highly concentrated within a small scientific elite, the scientific elite is itself a highly interconnected cluster of research interdependent influentials.

To investigate the possible conditions explaining a scientist's propensity for winning multiple prizes, we regressed the number of prizes a scientist wins, conditional on having won one prize, on variables measuring their genealogical and co-authorship ties (40) as specified above, individual talent using (41, 42) measures of productivity and H-index, and control variables for graduation year, university rank (43), and discipline using a Poisson regression with robust standard errors for scientist's graduation year as implemented in STATA.

Table 1 shows the regression results for genealogical predictors and controls and co-authorship variables and controls separately, and for the full model specification.

Controlling for other factors, both genealogical and collaboration networks significantly predict prizewinning. As expected, a scientist with a prizewinning advisor has a significantly higher propensity for becoming a multi-prize winner than a scientist that does not have a prizewinning advisor. Also, consistent with expectations, we find that the greater the number co-authorship relationships, the greater the propensity for becoming a multi-prizewinner. Contrary to our expectations, having a prizewinning co-author diminishes the probability of winning multiple prizes. One explanation for this effect is that a prizewinning co-author is deemed the driving force behind the scientific discoveries, which in turn reduces the perceived credit given to co-authors, a finding broadly consistent with the Mathew Effect (44).

**Discussion**

The motivations to conduct outstanding scientific work and the impact that work may have on science is arguably driven by both scientists' desire to know and to be recognized by their peers. In this work, we used new available large scale data on prizes and their winners to quantify the proliferation of prizes, reveal the hidden connections prizes and prizewinner make within and across disciplines, and explain the increasing concentration of scientific impact and fame within a proportionately smaller and smaller and more tightly interconnected elite.

Several areas of future work follow from our analyses. First, while we discovered the explosive growth in scientific prizes, we did not explain why it has occurred or its varied rate of growth across different subfields of science (45). Presumably, prizes grow to represent a broader range of ideas and specializations. However, we showed that while more scientists win prizes each year, it is also true that multiple prizes are increasingly won by a smaller and smaller set of winners. The increasing concentration would seem to work at cross purposes with the inclusiveness and inequality ethos of science. Thus, future research might delve more deeply into the nature of prizes to see if the proliferation of prizes is meeting their goals.

The arguably less optimistic finding of this research is that while prizes have increased in number the set of scientific influentials is increasingly made up of a smaller and more intertwined set of scientists. These scientists are linked through genealogical and co-authorship networks. If these networks operate like other social networks, they may provide continuous learning opportunities, better divisions of specialized labor, and inspiration for ideas for scientists but may also be vulnerable to in-group thinking that can keep good ideas out (34, 46). Further work on how these networks form and are related to the novelty and originality of work could help explain which processes are dominant and when in a science.

A final area of future research is to examine how prize interlocks – prizes that are often won by the same scientists – influence knowledge transfer. In the paper, we drew on research that indicated that such interlocks in a network perform the function of knowledge pathways between different regions of the network. It would seem plausible that a similar function is performed by interlocks in this network – transferring knowledge from area of discipline to another or longer distances between disciplines. For example, Daniel Kahneman, a psychologist by training and multi-prizewinner in psychology before winning a Nobel in economics has noted that the Nobel in economics legitimated and diffused his ideas about non-rational decision-making in economics, leading to one of his psychology papers becoming one of the most cited papers in economics (47). If one could development appropriate measures to track systematically knowledge transfers through interlocks, they could conceivably predict scientific trends, novel combinations of ideas, and possibly the next big thing in science.

## Materials and Methods

### Data sources

We merged three large datasets together: (i) prizewinners, (ii) scientific genealogy, (iii) publications. We parsed and curated the prizewinner data from the open web sources including Wikipedia and the prize official websites such as the www.nobelprize.org/. These contain information on each winner's name, prize name, prize year(s), gender, birth place, and career institutional affiliation(s). The scientific genealogy data were acquired from the Academic Family Tree (AFT) (https://neurotree.org) and the Mathematics Genealogy Project (www.genealogy.math.ndsu.nodak.edu). In these dataset, advisors and their students are listed. From the Web of Science dataset, we extracted publication and collaboration for every scholar in the Academic Family Tree datasets. The WoS contains more than 60M papers published from 1900 to 2015.

### Prize Prestige Scores

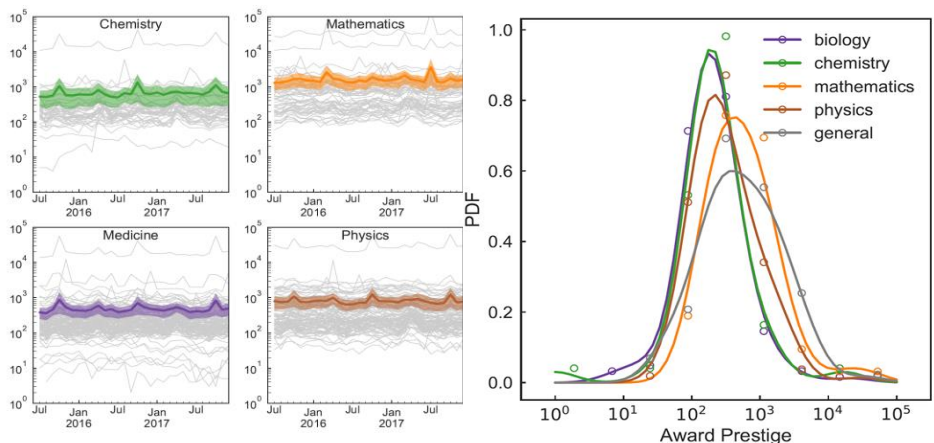

Wikipedia Pageviews by the Pageview API provide monthly pageviews of each prize from July 1, 2015 to December 31, 2017. The following figure shows the change in page views for each prize in biology, chemistry, mathematics, and physics and the distribution of prestige for all prizes in four primary disciplines: math, physics, chemistry and biology.

**Prize Network**

The prize network was constructed using 307 leading prizes and 6,061 weighted links between prizes. In the prize network, prizes are nodes, and links between two prizes *i* and *j* are formed when a least one scientist has won both prizes. In the main paper, we include a bipartite projection of the prize network in Fig. 3. In Fig 3., to provide an informative view of the prize network, we consider prizes with at least 10 recorded winners and winners with fewer than 10 different awards during their careers to reduce the visual density of the network, and Fig. 3 shows only the critical links identified by the principal link extracting algorithm (48, 49) for *p*-value equal to 0.3.

**Genealogical and Collaboration Networks**

The network of scientific winners is constructed using scientists' genealogical and co-authorship information. The genealogy structure for scientists is an acyclic tree, and each link in the network represents a genealogy (between advisors and their PhD students or postdocs) relationship. The co-authorship network may have a weighted link proportional to the number of papers two scientists coauthored. We combined the two networks into one containing 2,034 winners, circa 1960-2017. 1,504 (74%) of the winners are connected in a giant component.

**Acknowledgment:** We thank the Clarivate Analytics (the owner of the Web of Science database), the Mathematics Genealogy Project, Academic Family Tree, Wikipedia and Wikidata, for making their data available for analysis. **Funding:** This material is based upon work supported by, or in part by, the U. S. Army Research Laboratory and the U. S. Army Research Office under grant number W911NF-15-1-0577, QUANTA: Quantitative Network-based Models of Adaptive Team Behavior, and support, Army Research Labs (#W911NF-09-2-0053), the NIH (#R01GM112938), Kellogg School of Management, and the Northwestern Institution on Complex Systems (NICO) for generous funding for this project. **Author Contributions:** YM collected data, YM and BU designed, conducted analyses, and wrote the paper. **Competing interests:** The authors declare no competing interests in this work.

**Fig. 1 A century of scientific prizes.** Panels A and B represent the relative proliferation of scientific prizes (A) and prizewinners (B) from 1900 to 2015. Panel A shows the proliferation of prizes and the proliferation of separate scientific disciplines. Before 1980, there is a similar proliferation rate for disciplines and prizes, though there were nearly twice as many scientific disciplines as prizes; After 1980 prizes proliferate continue to proliferate at the pre-1980 rate and by 2015 outnumber the number of scientific fields at a 2:1 ratio.

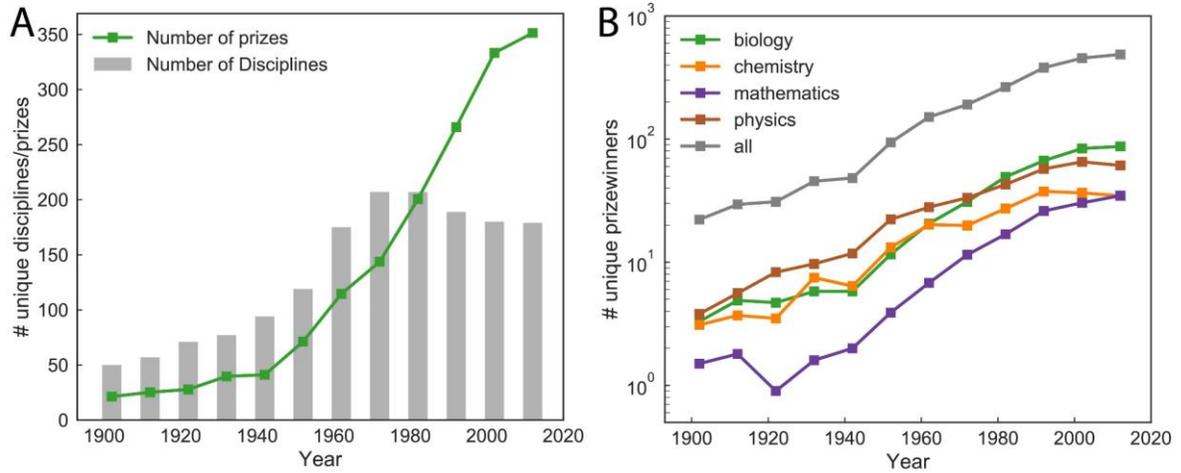

**Fig. 2 The exponential distribution of scientific prizewinning.** Plot and inset show number of prizes per prizewinner and the change in the distribution of prizes per winner before and after 1985, the midpoint of our data. The distribution on prizes per prizewinner fits an exponential distribution and indicates that many different prizes are won by a relatively small number of scholars. For example, over 60% of prizewinners are two-time winners of different prizes and about 15% are 5 time winners of different prizes. The inset shows that this heavy concentration of diverse prizes among a relatively small scientific elite has intensified. Despite there being nearly twice as many prizes after 1985 than before 1985, fewer scientists win a larger share of available prizes.

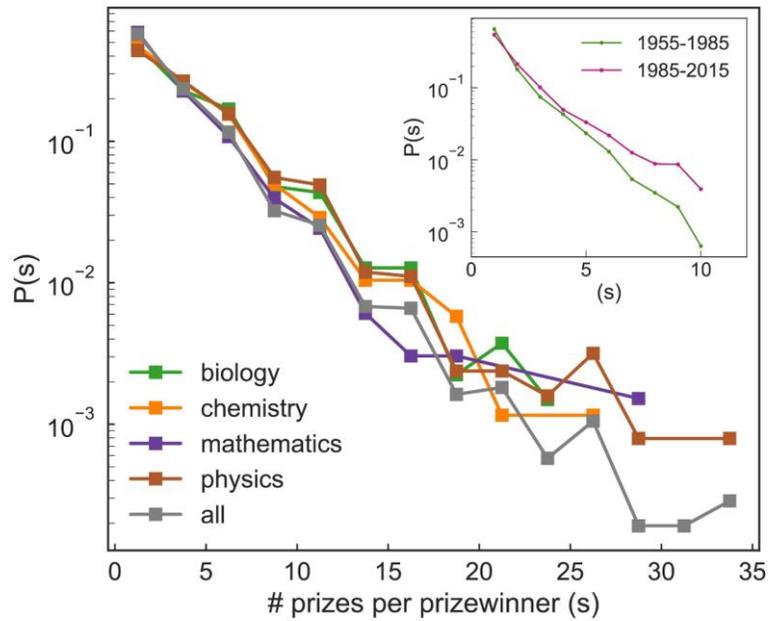

**Fig. 3 The scientific prize network.** In the network, nodes denote prizes, node size reflects the prize's relative prestige within its discipline, and links form between a pair of prizes when the same scientist wins both prizes. The link weight is proportional to the number of scientists who won both prizes. The network is stratified by prestige of prizes and clustering within disciplines and among some prizes with certain prizes playing an important role in creating knowledge interlocks between disciplines.

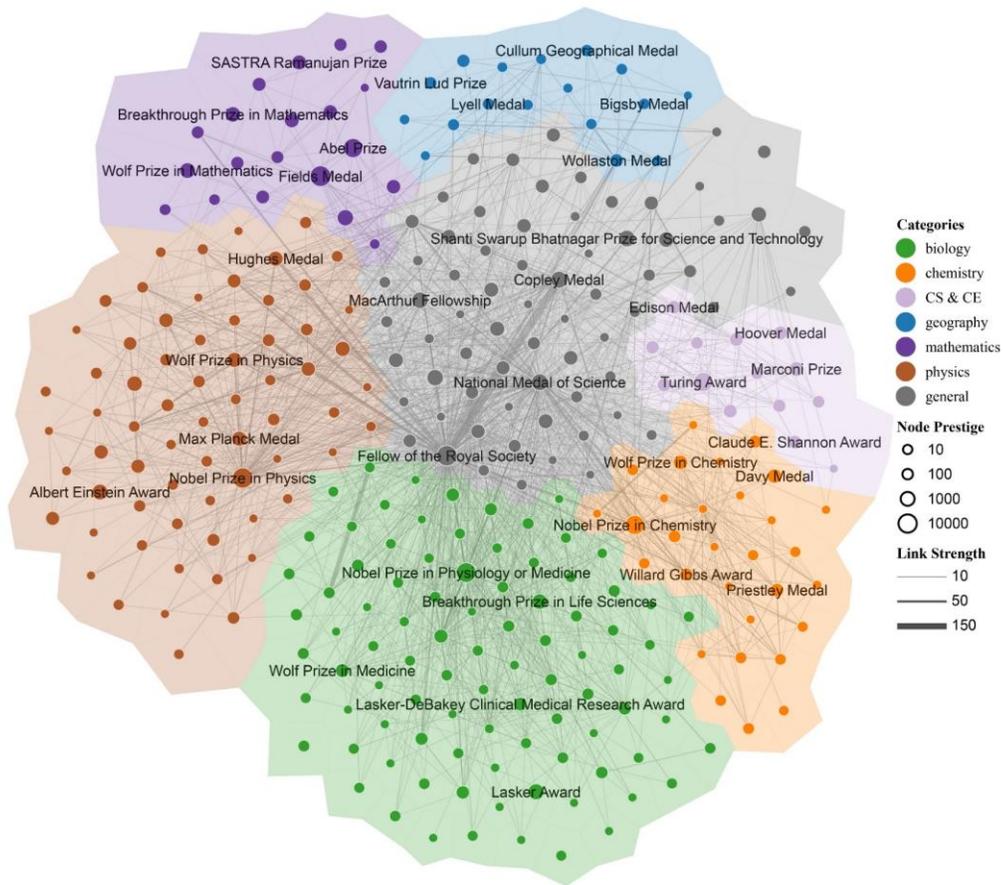

**Fig. 4 Scientific prize transition matrix.** When two prizes are won by the same scholar(s) they form an interlock between prizes indicate the pathway knowledge follows within and between disciplines as well as a scholar propensity for winning a prize given they have won another prize. The modular structure of the transition matrix reflects the modularity of the prize network and shows that disciplines vary in the transition between prizes. Chemistry and physics have relatively high transition probabilities among prizes relative to math and biology. We observe also that math and physics share a relatively high number of prizewinners but chemistry and biology share relatively few with physics and math or with one another. General prizes play the unique role of integrating diverse sciences.

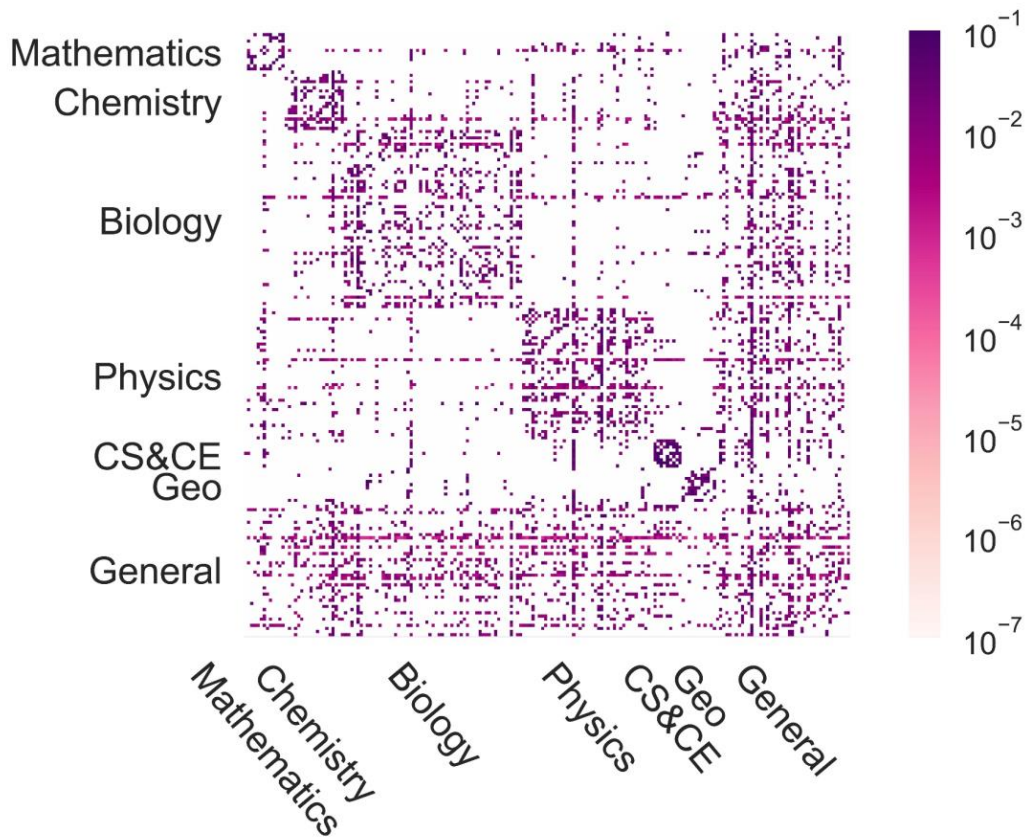

**Fig. 5 Social network of prizewinners.** Nodes (not shown) represent prizewinners and links represent the presence of genealogical or collaborative relationships between winners. Network shown here contains 830 winners, in which only strong co-authorship ties (winners coauthored more than 3 papers together) are shown. Genealogical relationships, which are a scholar's primary, formative relationships, are more evenly distributed throughout the network than are co-authorship ties, which are noticeably concentrated in the dense center of the network and continue to grow in number throughout most scholar's careers. This pattern suggests that genealogy plays a critical role in getting into the network but that co-authorship ties become increasingly important to prizewinning as a scientist's career advances.

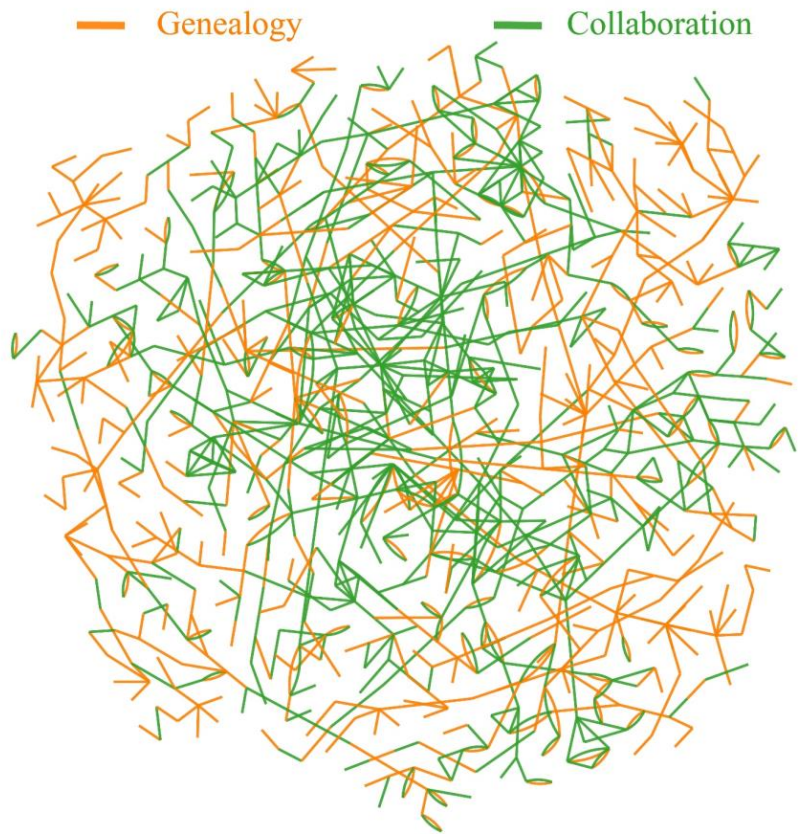

Table 1: **Poisson Regression Estimates of the Predicted Number of Prizes Won by a Multi-Prizewinning Scientist, 1960-2015.** The table reports estimates from a Poisson regression that regresses a scientist's propensity of winning multiple prizes conditional on having won a prize based on a genealogical and co-authorship networks, individual human capital variables, and controls for discipline, university prestige, and graduation date.

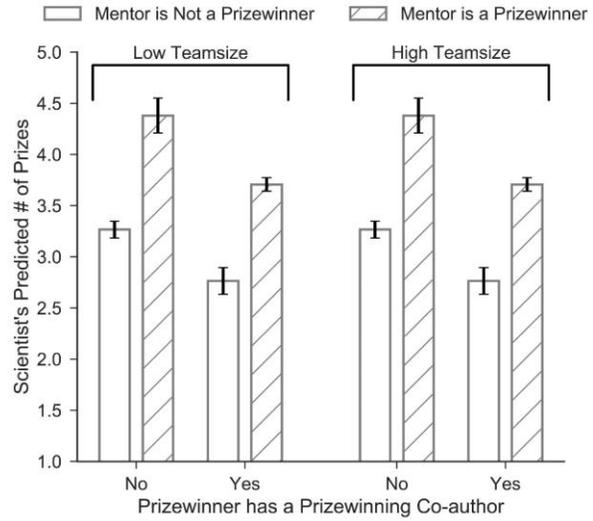

|  | (1) | (2) | (3) |
|---|---|---|---|
| Mentor is Prizewinner (Y/N) | 0.114*** (-0.04) |  | 0.111** (-0.04) |
| Co-author is Prizewinner (Y/N) |  | -0.187*** (-0.04) | -0.185*** (-0.04) |
| # Collaborators (Pre-prize) | 0.090** (-0.03) | 0.103** (-0.03) | 0.106** (-0.04) |
| Avg. teamsize | 0.008*** (-0.0006) | 0.007*** (-0.0005) | 0.008*** (-0.0004) |
| # Publications (Pre-prize) | -0.032 (-0.05) | -0.021 (-0.05) | -0.018 (-0.05) |
| H-Index (Pre-Prize) | 0.009*** (-0.003) | 0.010*** (-0.003) | 0.010*** (-0.003) |
| University prestige | √ | √ | √ |
| Discipline | √ | √ | √ |
| Graduation year | √ | √ | √ |
| Constant | 0.488*** (-0.11) | 0.507*** (-0.12) | 0.416** (-0.13) |
| BIC | 3057 | 3049 | 3043 |
| Observations | 639 | 639 | 639 |

** $p<0.01$, *** $p<0.001$